\documentclass[conference]{IEEEtran}
\usepackage{amsmath,amssymb,bm}
\usepackage{graphicx}
\usepackage{color}
\usepackage{subeqnarray}

\graphicspath{{./prog_EH/fig/}}

\newcommand{\qh}{{\bf h}}

\newcommand{\qp}{{\bf p}}

\newcommand{\qt}{{\bf t}}

\newcommand{\qv}{{\bf v}}
\newcommand{\qw}{{\bf w}}

\newcommand{\qz}{{\bf z}}

\newcommand{\be}{\begin{equation}} \newcommand{\ee}{\end{equation}}
\newcommand{\bea}{\begin{eqnarray}} \newcommand{\eea}{\end{eqnarray}}

\begin{document}

\title{Exploring Green Interference Power for Wireless Information and Energy Transfer in the MISO Downlink}
\author{\IEEEauthorblockN{Gan Zheng\IEEEauthorrefmark{1}, Christos~Masouros\IEEEauthorrefmark{2}, Ioannis Krikidis\IEEEauthorrefmark{3}, Stelios Timotheou\IEEEauthorrefmark{4}}
\IEEEauthorblockA{\IEEEauthorrefmark{1} School of Computer Science
and Electronic Engineering, University of Essex, Colchester, UK, CO4
3SQ} \IEEEauthorblockA{\IEEEauthorrefmark{2} Department of
Electronic \& Electrical Engineering, University College London,
Torrington Place, London, WC1E 7JE }
   \IEEEauthorblockA{\IEEEauthorrefmark{3} Department of Electrical
and Computer Engineering,  University of Cyprus, Cyprus}
\IEEEauthorblockA{\IEEEauthorrefmark{4} KIOS
  Research Center for Intelligent Systems and Networks, University of Cyprus, Cyprus}
 Email: ganzheng@essex.ac.uk, chris.masouros@ieee.org,
krikidis@ucy.ac.cy, timotheou.stelios@ucy.ac.cy}

\maketitle

\begin{abstract}
In this paper we propose a power-efficient transfer of information
and energy, where we exploit the constructive part of wireless
interference as a source of green useful signal power. Rather than
suppressing interference as in conventional schemes, we take
advantage of  constructive interference among users, inherent in the
 downlink, as a source of both useful information and wireless
energy. Specifically, we propose  a new precoding design that
minimizes the transmit power while guaranteeing the quality of
service (QoS) and energy harvesting constraints for generic phase
shift keying  modulated signals. The QoS constraints are modified to
accommodate constructive interference. We derive   a sub-optimal
solution and a local optimum solution to the precoding optimization
problem.
  The proposed precoding reduces the transmit
power compared to conventional schemes, by adapting the  constraints
to accommodate constructive interference as a source of useful
signal power.  Our simulation results show significant power savings
with the proposed data-aided precoding compared to the conventional
precoding.
 \end{abstract}

\section{Introduction}
Recently simultaneous wireless information and energy transfer
 via the radio frequency energy harvesting (EH)
technology has emerged as a new solution for sustainable wireless
network operation. In a multiuser   scenario, interference signals
provide a source for energy harvesting.
On the other hand, there have been some efforts in exploiting
constructive interference (CI) to improve the users'
quality-of-service (QoS)  and save the transmit power. This paper
aims to exploit both data and channel state information (CSI) at the
transmitter to facilitate the information and energy transfer in a
multiple-input single-output (MISO) broadcast channel. Below we give
brief literature review on simultaneous wireless information and
energy transfer and data-aided precoding.

The fundamental concept of simultaneous wireless transmission of
energy and information is introduced in \cite{GROV} from an
information theoretic standpoint.
In \cite{ZHA} the authors   discuss two practical receiver
structures for simultaneous wireless
 information and energy transfer termed as   ``time switching'' (TS)
and   ``power splitting'' (PS), to separate the received signal  for
decoding information and harvesting energy, respectively.
Furthermore, multi-antenna technology can greatly facilitate the
information and energy transfer. The optimal precoding design for
energy and information transfer in a  MISO broadcast channel is
studied in \cite{Zhang-14}.   The achievable information and energy
transmission trade-offs  are investigated in \cite{Clerckx-13} for a
2-user multiple-input multiple-output (MIMO) interference channel
(IC), based on different combinations of receiver modes. Joint
information and energy transfer  is studied for a general K-user
MISO IC  in \cite{Zheng} based on PS receivers.

Regarding the precoding design for a broadcast channel, the
conventional view of data  is that they are totally random and
independent, and precoding design is independent of data. The
proposed precoding in this paper is based on the viewpoint that data
could be incorporated into the precoding design to better exploit
the interference, which is first introduced in \cite{SCI,CR, ComMag}
where analytical interference classification criteria and
low-complexity precoders based on channel inversion where derived.
The analysis showed with the knowledge of both CSI and data at the
base station (BS),  some interfering data could contribute to the
detection of the desired symbol and thus are classified as CI.

In this work we aim to optimize data-aided precoding design in a
MISO broadcast channel with PS receivers  by exploiting constructive
interference as a useful source for both signal and energy transfer.
 We study the problem of transmit power minimization for guaranteeing
both the signal-to-interference-plus-noise ratio (SINR) and EH
constraints. The problem  has an interesting interpretation of
multicast formulation. We propose a sub-optimal solution and a local
optimum solution for the precoding design. Our results show that
compared to the conventional precoding, the proposed data-aided
precoding leads to 5-7 dB of power saving in high signal-to-noise
ratio (SNR) region.

\emph{Notation:} We use the upper case boldface letters for matrices
and lower case boldface letters for vectors. $(\cdot)^{*}$ and
$(\cdot)^{T}$ denote  the conjugate   and transpose, respectively.
   $\|\cdot\|$ stands for  the Frobenius norm.
 A  complex Gaussian random vector variable $\qz$ with mean $\bm\mu$
and variance variance $\bm\Sigma$ is represented as
$\qz\sim\mathcal{CN}\left(\bm\mu, \bm\Sigma\right)$.
$\mathbb{E}\{\cdot\}$ denotes the expectation.

\section{System Model}
 Consider a   Gaussian MISO broadcast channel where an $N$-antenna BS
 transmits both signals and energy to $K$ single-antenna users. For user $i$, its channel vector,
 precoding vector, received noise, data, SINR and EH constraints are
 denoted as $\qh_i^T, \qt_i$,  $n_i$, $d_i$, $\Gamma_i$, $E_i$,  respectively. The    phase shift
keying (PSK)  modulated symbol   can be expressed as $d_i= d
e^{j\phi_i}$ where $d$
 denotes the constant amplitude  and $\phi_i$ is the
 phase. Without loss of generality, we assume $d=1$.
The total transmit power is
 \be\label{eqn:PT}
    P_T =  \mathbb{E}\left\{\left\|\sum_{k=1}^K \qt_k
    d_k\right\|^2\right\}.
 \ee
 $n_i\sim \mathcal{CN}(0, N_0)$ is the additive white Gaussian noise (AWGN).
All wireless links exhibit independent frequency non-selective
Rayleigh block  fading.     The received signal at user $i$ is
 \bea\label{eqn:rs}
    y_i &=& {\qh_i^T} \sum_{k=1}^K \qt_k d_k + n_i.
 \eea
To decode the information and harvest RF energy at the receiver
side, the practical PS technique  \cite{ZHA} is used. Specifically,
the receiver splits the RF signal into two parts: one for
information decoding and the other for energy  harvesting, with
relative power of $\rho_i$ and $1-\rho_i$, respectively.

The signal for information decoding is expressed as %
\bea\label{eqn:sig:ID} \tilde y_{i} &=& \sqrt{\rho_i} {y}_i + \tilde n_i\nonumber\\
&=& \sqrt{\rho_i}  {\qh_i^T} \sum_{k=1}^K \qt_k  d_k +
\sqrt{\rho_i}n_i  + \tilde n_i,    \label{sigDecoder} \eea where
$\tilde n_i\sim \mathcal{CN}(0, N_C)$ is the complex AWGN introduced
in the RF to baseband conversion in the decoding process, which is
independent of $n_i$.

The signal for energy harvesting  is   \be\label{eqn:sig:EH}
    \bar y_{i} = \sqrt{1-\rho_i} {y}_i = \sqrt{1-\rho_i} \left( {\bf h}_{i}^T\sum_{k=1}^K{\bf t}_k d_k+n_i\right)
\ee with average power \bea \label{eqn:P:EH}
    P_{i}& =& (1-\rho_i)    \mathbb{E}\left\{\left|{\qh_i^T} \sum_{k=1}^K \qt_k d_k + n_i \right|^2\right\}
    \nonumber.
\eea
 The problem of interest is to minimize the
total transmit  power $P_T$ in \eqref{eqn:PT} subject to QoS (i.e.,
SINR) constraints $\{\gamma_i\}$ and energy harvesting constraints
$\{E_i\}$, respectively.
 This will be achieved by optimizing beamforming design, power
allocation and splitting, as well as CI.

In the following, we first review the conventional precoding design
then we introduce the proposed approach based on constructive
interference.
\section{Review: Conventional  Precoding} 
 In conventional MISO downlink precoding, users' data are independent to each other, i.e., $\mathbb{E}
(d_i^* d_j)=0, \forall j\ne i$. In this case, the transmit power in
 \eqref{eqn:PT} becomes
 \be
    P_T = \sum_{i=1}^K \|\qt_i\|^2.
 \ee

 Based on the signal model \eqref{eqn:sig:ID} for information decoding, the received SINR for user $i$ is given by
\begin{equation}\label{eqn:SINR}
{\Gamma}_i^{con}= \frac{ |\qh_{i}^T\qt_i|^2}{\sum\limits_{j=1,j \ne
i}^K |\qh_{i}^T\qt_j|^2 + N_0 + \frac{N_C}{\rho_i}} .
\end{equation}

The harvested energy can be derived from \eqref{eqn:P:EH} as
\be\label{eqn:power}
    P_i^{con} = (1-\rho_i) \left(    \sum_{k=1}^K |{\qh_i^T} \qt_k|^2    + N_0\right).
\ee Consequently, the power minimization problem with both QoS and
EH constraints can be formulated as
\bea\label{eqn:prob:EH:conventional}
\min_{\{\qt_i,  \rho_i\}}  &&  \sum_{i=1}^K \|\qt_i\|^2  \\
\mbox{s.t.}&&
   \Gamma_i^{con}   \ge \gamma_i,\notag\\
&&  P_i^{con}\ge E_i,\notag\\
&&    0<\rho_i<1, \forall i.\label{eqn:energy} \notag\eea It is easy
to see that formulation \eqref{eqn:prob:EH:conventional} is
non-convex and hence   challenging to solve. In our previous work
\cite{Zheng}, we have used semidefinite programming relaxation to
tackle it and we show that the relaxation  is tight for 2-user and
3-user MISO IC. This result is extended for the general MISO
downlink in \cite{Zhang-14}.

\section{Proposed Precoding with constructive interference}
\subsection{Problem Formulation}
Traditional precoding approaches treat  users' data as totally
random and independent information streams, however, for the PSK
signalling considered, it is clear that each PSK symbol is merely a
rotated version of another, i.e., $d_j = d_i e^{j(\phi_j-\phi_i)}$,
therefore one user's data do not always generate harmful
interference to others. With the knowledge of both the instantaneous
CSI and the data symbols  at the BS, the received interference can
be classified to be constructive or destructive.
In brief, while destructive interference  performance, CI moves the
received symbols away from the decision thresholds of the
constellation and thus improves the detection. We refer the readers
to \cite{SCI, CR, ComMag} for further details. The main idea of the
proposed precoding is to exploit the CI for both information
decoding and energy harvesting.

The received signal  at user $i$ in \eqref{eqn:rs} can be rewritten
as
 \bea
    y_i &=& {\qh_i^T} \sum_{k=1}^K \qt_k d_k + n_i\nonumber\\
      &=& {\qh_i^T} \sum_{k=1}^K \qt_k  e^{j(\phi_k-\phi_i)} d_i +
      n_i.
 \eea
The information decoding part can be written as
 \bea
    \sqrt{\rho_i} y_i + \tilde n_i = \sqrt{\rho_i} {\qh_i^T} \sum_{k=1}^K \qt_k  e^{j(\phi_k-\phi_i)} d_i +
      \sqrt{\rho_i}  n_i + \tilde n_i.
\eea

\begin{figure}[tb]
\centering
\includegraphics[width=0.55\textwidth]{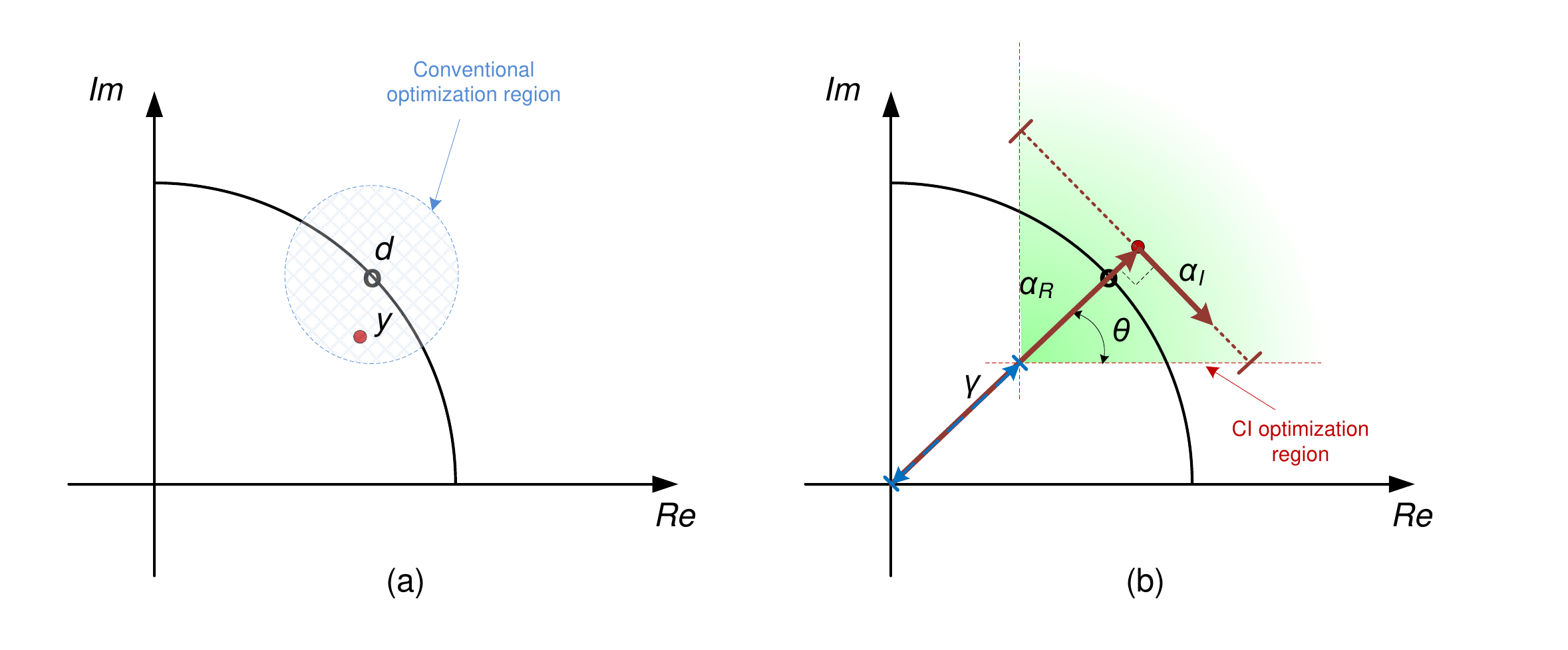}
  \caption{Illustration of constructive interference for information decoding, QPSK example.}
\label{FIGG11}
\end{figure}

We illustrate the derivation of the SINR constraint for the example
of QPSK in Fig. \ref{FIGG11}. Here, Fig. \ref{FIGG11}(a) represents the conventional optimization region and Fig. \ref{FIGG11}(b) shows the proposed optimization region. We have used the definitions
$\alpha_r=\mbox{Re} \left({\qh_i^T} \sum_{k=1}^K \qt_k
e^{j(\phi_k-\phi_i)}\right)$, $\alpha_i=\mbox{Im} \left({\qh_i^T}
\sum_{k=1}^K \qt_k e^{j(\phi_k-\phi_i)}\right)$ and
$\gamma=\sqrt{\Gamma_i \left(N_0+\frac{N_C}{\rho_i}\right)}$, where
$\mbox{Re}(x)$ and $\mbox{Im}(x)$ denote the real part and imaginary
part of $x$, respectively. Clearly, $\alpha_r$ represents the
amplification of the received constellation point due to the CI and
$\alpha_i$ gives the angle shift from the original constellation
point.
 In conventional precoding optimization,   $\alpha_r$ and
$\alpha_i$ are constrained such that the received symbol is
contained within a circle (denoted by the dashed circle)   around
the nominal constellation point, so that the interference caused by
the other symbols is limited.  In contrast to this, the concept of
CI is exploited to   allow a relaxation of $\alpha_r$ and $\alpha_i$
for all transmit symbols, under the condition that the interference
caused is constructive, lying in the green shaded sector in the
diagram \cite{CR}. It can be seen that $\alpha_r$ and $\alpha_i$ are
allowed to grow infinitely, as long as their ratio is such that the
received symbol is contained within the constructive area of the
constellation, i.e. the distances from the decision thresholds, as
set by the SNR constraints $\gamma$, are not violated.  It can be
seen that the angle of interference need not to be strictly aligned
with the angle of the useful signal, as long as it falls within the
constructive area of the constellation. For a given modulation order
$M$ the maximum angle shift in the CI area is given by $\theta=
\pi/M$.  Using basic geometry we arrive at the SINR constraint
expressed as
 \be
  |\alpha_i| \le (\alpha_r-\gamma) \tan\theta,
  \ee
  which is expanded as
{\small \bea
&& \left|\mbox{Im} \left({\qh_i^T}
\sum_{k=1}^K \qt_k e^{j(\phi_k-\phi_i)}\right)\right|\le\label{eqn:SINR:DA}\\
&&\left(\mbox{Re} \left({\qh_i^T} \sum_{k=1}^K \qt_k
e^{j(\phi_k-\phi_i)}\right)- \sqrt{\Gamma_i \left( N_0+
\frac{N_C}{\rho_i}\right)} \right)\tan\theta.\notag \eea}

The harvested energy and the total transmit power can be derived
based on  \eqref{eqn:PT} and \eqref{eqn:P:EH}, respectively, as
$P_i=|{\qh_i^T} \sum_{k=1}^K \qt_k  e^{j(\phi_k-\phi_i)}|^2$ and $
P_T = \left\|\sum_{k=1}^K \qt_k  e^{j(\phi_k-\phi_i)}\right\|^2$.
Therefore, the power minimization problem subjective to both SINR
and EH constraints with the aid of  the CI can be formulated as
  \bea\label{eqn:QoS:EH0}
\min_{\{\qt_i,  \rho_k\}}&&     \left\|\sum_{k=1}^K \qt_k  e^{j(\phi_k-\phi_i)}\right\|^2 \\
 \mbox{s.t.}&&   \eqref{eqn:SINR:DA},\notag\\
   && |{\qh_i^T} \sum_{k=1}^K \qt_k  e^{j(\phi_k-\phi_i)}|  \ge  \sqrt{\frac{E_i}{ 1-\rho_i }  }, \notag\\
&&    0<\rho_i<1, \forall i.\notag \eea
By defining   $\tilde \qh_i
= \qh_i e^{j(\phi_1-\phi_i)}$ and $\qw\triangleq \sum_{k=1}^K \qt_k
e^{j(\phi_k-\phi_1)}$,
we can write \eqref{eqn:QoS:EH0} equivalently as
 { \bea\label{eqn:QoS:EH}
\min_{\{\qw,  \rho_i\}}  &&  \|\qw\|^2  \\
\mbox{s.t.} && \left|\mbox{Im}
\left({\tilde\qh_i^T}\qw\right)\right|\le\nonumber
\\&&\left(\mbox{Re} \left({\tilde\qh_i^T}
\qw\right)-\sqrt{\Gamma_i   \left( N_0+ \frac{N_C}{\rho_k}\right)} \right)\tan\theta,\notag\\
   && |{\tilde\qh_i^T} \qw |^2  \ge   {\frac{E_i}{ 1-\rho_i } }, \label{eqn:cons30} \\
&&    0<\rho_i<1, \forall i. \nonumber \eea}
 Although the
reformulation \eqref{eqn:QoS:EH} seems a trivial step, it indicates
that the original broadcast channel now reduces to a virtual
multicast channel with common message to all users \cite{Sidir}.

The problem \eqref{eqn:QoS:EH} is   nontrivial to solve because of
the nonconvex constraint $|{\tilde\qh_i^T} \qw |^2  \ge {\frac{E_i}{
1-\rho_i } }$. The rest of this section is devoted to solving the
multicast problem \eqref{eqn:QoS:EH}.

\subsection{A Sub-optimal Solution}
 We first tackle the   nonconvex constraint $|{\tilde\qh_i^T} \qw |^2  \ge
{\frac{E_i}{ 1-\rho_i } }$  by adding a new constraint
$\mbox{Im}\left({\tilde\qh_i^T}\qw\right) =0$ then find a
sub-optimal solution. With this new constraint, the problem
\eqref{eqn:QoS:EH} becomes
 \bea\label{eqn:QoS:EH:theta0}
\min_{\{\qw,  \rho_i\}}  &&  \|\qw\|^2  \\
\mbox{s.t.} &&  \mbox{Im}\left({\tilde\qh_i^T}\qw\right) =0,\nonumber\\
&& \mbox{Re} \left({\tilde\qh_i^T} \qw\right) \ge
\sqrt{\Gamma_i   \left( N_0+ \frac{N_C}{\rho_i}\right)},  \label{eqn:cons2}\\
   && \mbox{Re} \left({\tilde\qh_i^T} \qw\right)  \ge   \sqrt{{\frac{E_i}{ 1-\rho_i } }},\label{eqn:cons3}\\
&&    0<\rho_i<1, \forall i.\nonumber \eea
 It is not difficult to
verify that  both $\sqrt{\Gamma_i   \left( N_0+
\frac{N_C}{\rho_i}\right)}$ and $ \sqrt{{\frac{E_i}{ 1-\rho_i } }}$
are  convex functions in $\rho_i$, so the above problem is convex.
However, it can be further simplified. Observe that at the optimum,
both constraints \eqref{eqn:cons2} and \eqref{eqn:cons3} can be
satisfied with equality, i.e.,
 \be
    \Gamma_i   \left( N_0+ \frac{N_C}{\rho_i}\right) = {{\frac{E_i}{ 1-\rho_i }
    }}, \forall i,
 \ee
from which we can solve
\be
    \rho_i^* = \frac{-B - \sqrt{B^2-4AC}}{2A},
\ee
 where $A =-\Gamma_i N_0, B = \Gamma_i N_0 - E_i - \Gamma_i N_C, C = \Gamma_i
 N_C$. $\rho_i^*$ is independent of $\qw$.
With $\rho^*$, the resulting problem is written as
\bea\label{eqn:QoS:EH:theta01}
\min_{\{\qw \}}  &&  \|\qw\|^2  \\
\mbox{s.t.} &&  \mbox{Im}\left({\tilde\qh_i^T}\qw\right) =0,\notag\\
&& \mbox{Re} \left({\tilde\qh_i^T} \qw\right) \ge \sqrt{\Gamma_i
\left( N_0+ \frac{N_C}{\rho_i^*}\right)},\forall i  \nonumber. \eea
This is a quadratic problem with    linear constraints in $\qw$,
therefore easy to solve.
\subsection{The Proposed DC Algorithm}
 Here we proposed another approach to tackle the  nonconvex constraint \eqref{eqn:cons30}  $|{\tilde\qh_i^T} \qw |^2  \ge
{\frac{E_i}{ 1-\rho_i } }$ using difference of convex optimization
(DC)
 and successive convexification. This method will produce a local
 optimum solution which is normally better than the above
 sub-optimal solution.

  To illustrate the idea, we define $f(\qv) \triangleq |v_1|^2 + |v_2|^2, v_1 \triangleq \mbox{Re}({\tilde\qh^T} \qw), v_2 \triangleq \mbox{Im}({\tilde\qh^T} \qw)$. For convenience, we have dropped the user index.
 Then the linear approximation of $f(\qw)$ around the point $\qp$ can be expressed as
 \bea
    f(\qv) &=& f(\qp) +  \left(\frac{\partial f(\qv) }{ \partial\qv}|_{\qv=\qp}\right)^T(\qv -
    \qp)\nonumber\\
     &=& f(\qp) +  2\qp^T (\qv -    \qp).
 \eea
 Based on the above approximation, we propose the following
 algorithm to solve the problem \eqref{eqn:QoS:EH}.

\underline{The Proposed DC Algorithm:}
 \begin{enumerate}
    \item Initialize $\qp^{0}$.
    \item At the iteration $k$, solve the problem
\bea\label{eqn:QoS:EH:convex}
&& \min_{\{\qw,  \rho_i\}}  ~~  \|\qw\|^2  \\
&&\mbox{s.t.} ~~ \left|\mbox{Im}
\left({\tilde\qh_i^T}\qw\right)\right|\nonumber \\
&&\le\left(\mbox{Re} \left({\tilde\qh_i^T} \qw\right)-
\sqrt{\Gamma_i  (N_0 + \frac{N_C}{\rho_i}})\right)\tan\theta,\label{eqn:cons1}\\
   && \|\qp_i^{(k)}\|^2 + 2\sum_{b=1}^2
   {[\qp_i^{(k)}]}_b( {[\qv_i]}_b-[{\qp^{(k)}}]_b)  \ge   {\frac{E_i}{ 1-\rho_i } },\nonumber\\
&&{[\qv_i]}_1=\mbox{Re} \left({\tilde\qh_i^T} \qw\right),
{[\qv_i]}_2=
\mbox{Im} \left({\tilde\qh_i^T}\qw\right),  \nonumber\\
&&    0<\rho_i<1,\forall i.\nonumber \eea
\item Update $\qp^{(k+1)}=[\mbox{Re} ({\tilde\qh_i^T} \qw); ~~ \mbox{Im}  ({\tilde\qh_i^T} \qw)]^T$ until convergence.
 \end{enumerate}
The problem \eqref{eqn:QoS:EH:convex} is recognized as a convex
problem and can be optimally solved. The proposed DC algorithm is
proved to converge to a local optimum  and it has been widely used
in signal processing for communications \cite{DC_application}.

There is one practical difficulty remaining. In the constraint
\eqref{eqn:cons1}, although   $\sqrt{\Gamma_i  (N_0 +
\frac{N_C}{\rho_i})}$ is a convex function in $\rho_i$, it does not
follow any composition rule therefore it is not recognized to be
convex by most numerical solvers such as CVX \cite{cvx}. In order to
use CVX to solve it, we need the following procedures to explicitly
show that the constraint $\sqrt{N_0 + \frac{N_C}{\rho}}\le g$   is
convex where $g$ is a variable. For convenience, we have dropped the
user index and the constant $\Gamma_i$. The key step is to introduce
a new variable $u$ and rewrite  the constraint $\sqrt{N_0 +
\frac{N_C}{\rho}}\le g$ as \be\label{eqn:1} \left\{
\begin{array}{l}
    \sqrt{N_0 + u^2}\le g, \\
    \frac{N_C}{\rho}\le u^2 \Leftrightarrow
\frac{\sqrt{N_C}}{u}\le     \sqrt{\rho}.
\end{array}\right.
\ee
  It is easy to check that both of the above constraints in
  \eqref{eqn:1}  are   convex jointly in $(\rho, u, g)$ using the
  composition rules \cite{Boyd}, thus they can all be recognized by CVX.

  \subsection{Finding a Feasible Solution}
 The remaining issue is how to find a feasible solution for the proposed DC algorithm. This can be
 achieved by solving the problem \eqref{eqn:QoS:EH} without the EH
 constraint (15) for arbitrary $\rho_i  ~(0<\rho_i<1)$, which is a second-order cone program
  problem and can be optimally and efficiently solved.
If the obtained solution satisfies the EH constraint
$|{\tilde\qh_i^T} \qw |^2  \ge {\frac{E_i}{ 1-\rho_i } }$, then a
feasible solution is found; otherwise, increase the power $\qw$ such
that the constraint $|{\tilde\qh_i^T} \qw |^2  \ge {\frac{E_i}{
1-\rho_i } }, \forall i$, is satisfied. This power amplification
will not affect the SINR constraint.

\section{Numerical Results}
In this section, we numerically assess the performance of the
proposed data-aided CI precoding scheme. Channels are independent
and experience frequency flat Rayleigh fading with zero mean and
unit variance. We compare the required BS transmit power of the
proposed DC solution, the sub-optimal solution, and the conventional
solution without considering the CI described in Section III    .
Unless otherwise specified, QPSK is the default modulation scheme
and all users have the same SINR and EH constraints. $K=N_t=4$ and
$N_0=N_1=1$ are assumed.  All figures are produced by averaging the
 results of 100 channel instances.

\begin{figure}[h]
\centering
\includegraphics[width=2.6in]{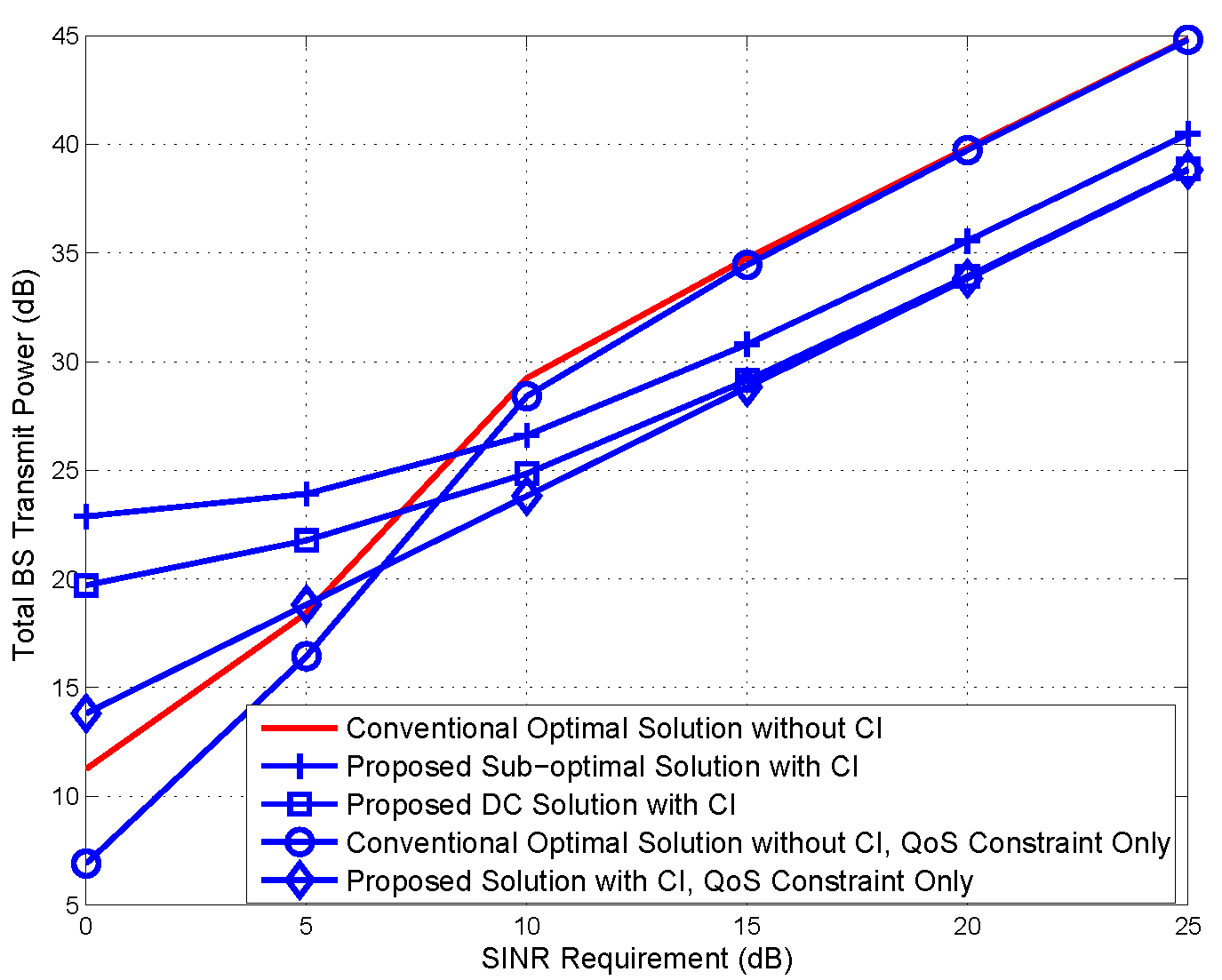}
  \caption{Transmit power vs. SINR constraints,  $E_i=10$ dB.}
\label{FIGG2}
\end{figure}

 Fig. \ref{FIGG2} depicts the    transmit power of the different
schemes versus the SINR requirement when the minimum required EH is
10 dB. The transmit power with only SINR constraints (no EH
constraints) for the conventional and CI based schemes is also shown
for comparison. We can see that when the SINR constraint is below 5
dB, the conventional solution outperforms the proposed CI based
solution. However, when the SINR constraint is greater than 10 dB,
there is a sharp rise of the transmit power for the conventional
scheme;   the proposed DC solution and the   sub-optimal solution
achieve power reduction of 7 dB and 5 dB, respectively. This can be
explained by the fact that when SINR is low, the required transmit
power is also low. Consequently, there is not sufficient
interference among users to exploit, which is also disadvantageous
for the proposed CI scheme. It is also seen that as SINR increases,
the transmit power converges to the solution with only SINR
constraint and this is because in the high SINR region, there is
sufficiently high interference thus EH constraints are automatically
satisfied.

\begin{figure}[h]
\centering
\includegraphics[width=2.6in]{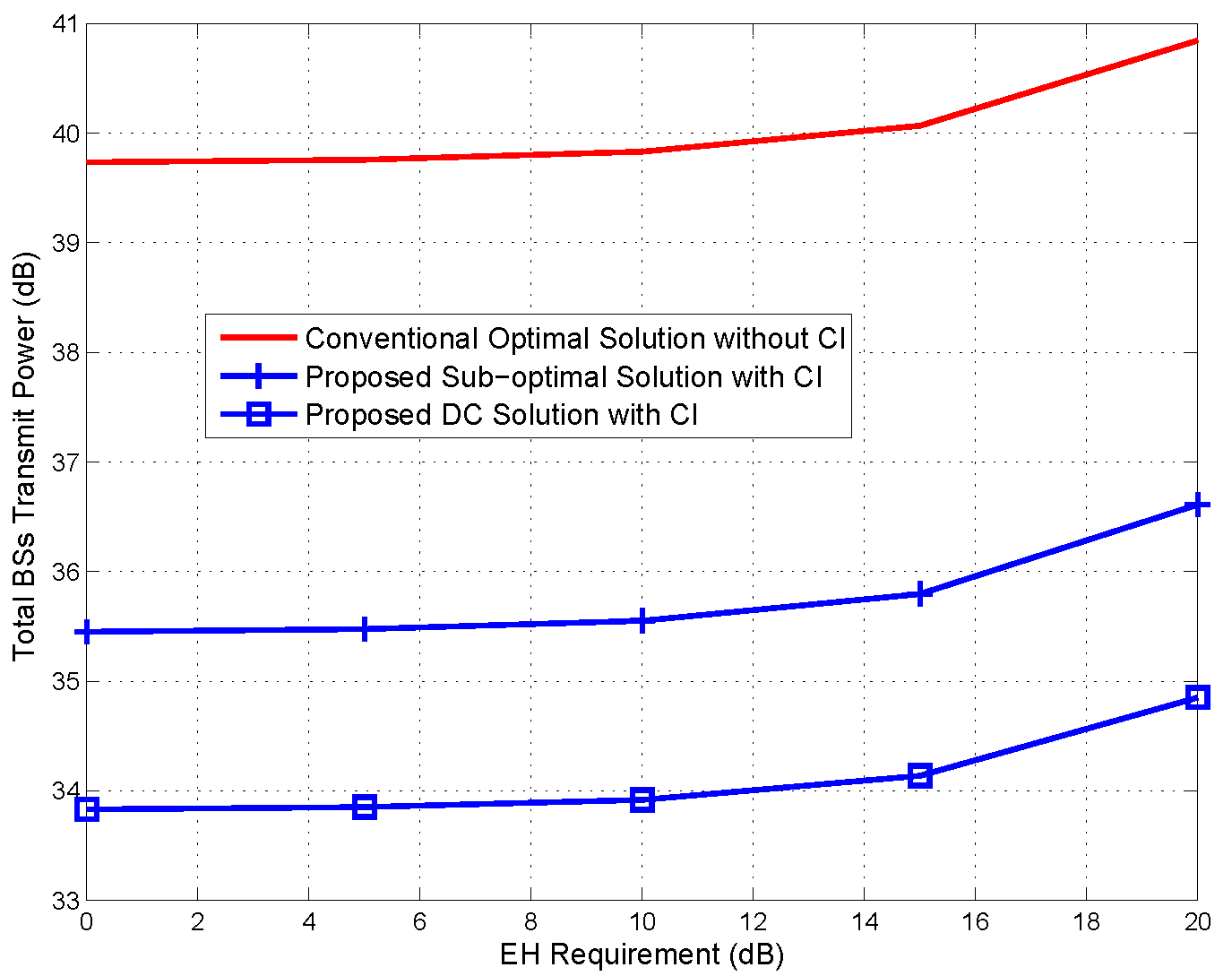}
  \caption{Transmit power vs. EH constraint constraints,  $\Gamma_i=20$ dB.}
\label{FIGG3}
\end{figure}

In Fig. \ref{FIGG3}, we compare the transmit power against the EH
requirement when the minimum required SINR is 20 dB. In this case,
the interference among users is high and therefore the proposed CI
based solutions outperform the conventional solution. Similar
performance gain in terms of power reduction as Fig. \ref{FIGG2} is
observed consistently in the whole EH region.

\begin{figure}[h]
\centering
\includegraphics[width=2.6in]{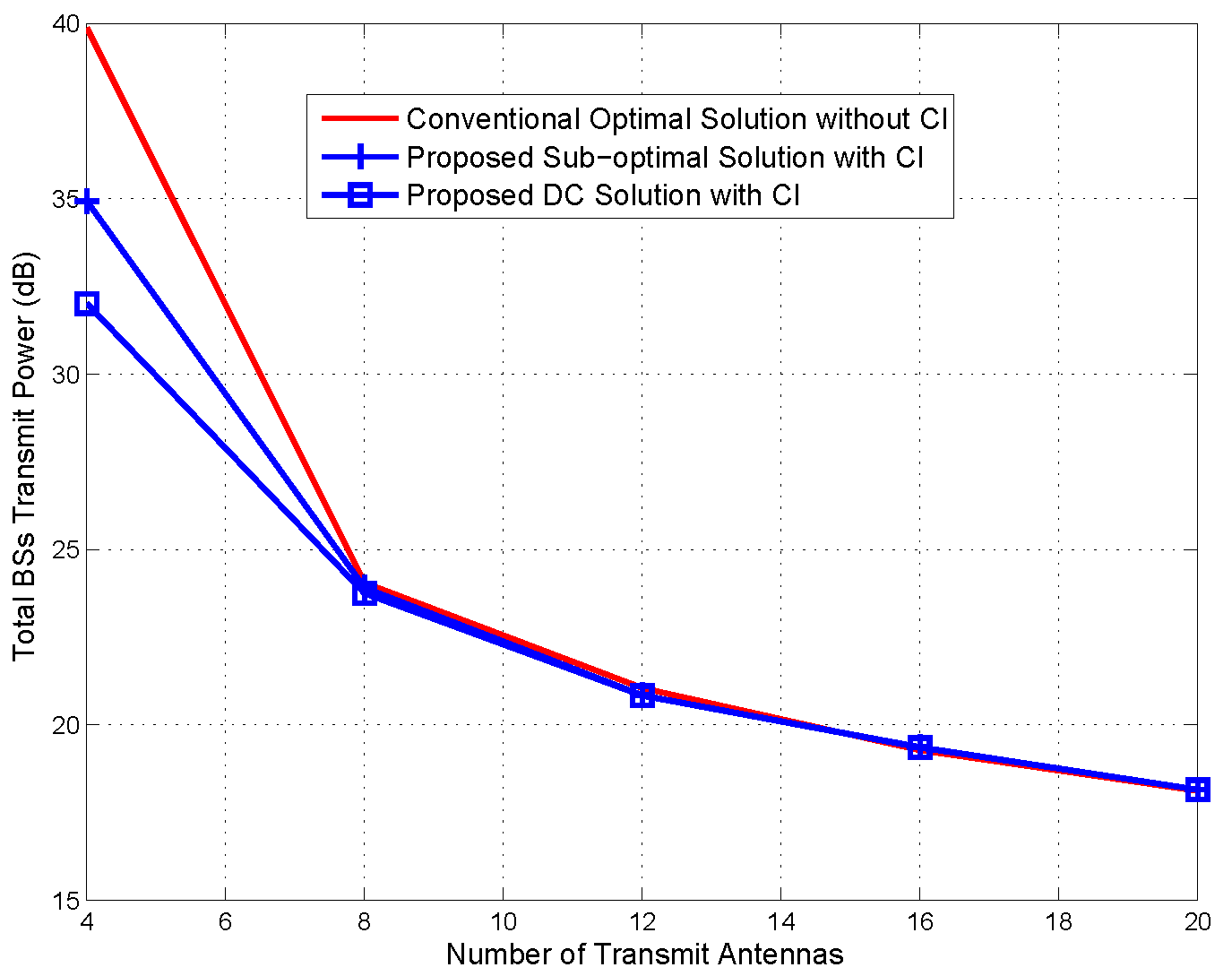}
  \caption{The effect of the number of transmit antennas,  $\Gamma_i=E_i=20$ dB.}
\label{FIGG4}
\end{figure}

Finally the effect of the number of transmit antennas is illustrated
in Fig. \ref{FIGG4}, when both SINR and EH constraints are 20 dB. As
can be seen, the benefit of the proposed CI solution diminishes
compared to the conventional scheme as the number of transmit
antennas increases. When the number of transmit antennas is greater
than 8, the performance gap among different schemes is negligible.
One possible reason for this observation is that when the number of
transmit antennas is large, users' channels tend to be more
orthogonal to each other  resulting in low interference.

\section{Conclusions}
 In this paper, we exploited the interference in MISO downlink to
 boost the performance for both information transfer and energy
 harvesting.   Both sub-optimal and DC based
 solutions are derived to achieve the precoder design.
Numerical results demonstrate that the proposed schemes
significantly reduce transmit power by  5-7 dB compared to the
conventional precoding design in the high SNR region. 

\end{document}